\documentclass[aps,prl,showpacs,preprintnumbers,twocolumn,superscriptaddress]{revtex4-2}
\usepackage{amsmath,amssymb}
\usepackage{bm}
\usepackage{tipa}
\usepackage{upgreek}
\usepackage{comment}
\usepackage{subfig}
\usepackage{mathrsfs}
\usepackage{graphicx}
\usepackage{braket}
\usepackage{enumitem}
\usepackage{mathbbol}
\usepackage{booktabs}
\usepackage{gensymb}
\usepackage[normalem]{ulem}
\usepackage{color}
\usepackage[colorlinks,bookmarks=true,citecolor=blue,linkcolor=red,urlcolor=blue]{hyperref}
\usepackage{hyperref}
\renewcommand{\vec}[1]{\mathbf{#1}}
\usepackage{pifont}
\newcommand{\cmark}{\ding{51}}
\newcommand{\xmark}{\ding{55}}

\begin{document}

	\title{Global Phase Diagram of the Normal State of  Twisted Bilayer Graphene} 
	
	\author{Glenn Wagner}
	\affiliation{Rudolf Peierls Centre for Theoretical Physics, Parks Road, Oxford, OX1 3PU, UK}
	\affiliation{Department of Physics, University of Zurich, Winterthurerstrasse 190, 8057 Zurich, Switzerland}
	\author{Yves H. Kwan}
	\affiliation{Rudolf Peierls Centre for Theoretical Physics, Parks Road, Oxford, OX1 3PU, UK}
	\author{Nick Bultinck}
	\affiliation{Rudolf Peierls Centre for Theoretical Physics, Parks Road, Oxford, OX1 3PU, UK}
	\affiliation{Department of Physics, Ghent University, Krijgslaan 281, 9000 Gent, Belgium}
	\author{Steven H. Simon}
	\affiliation{Rudolf Peierls Centre for Theoretical Physics, Parks Road, Oxford, OX1 3PU, UK}
	\author{S.A. Parameswaran}
	\affiliation{Rudolf Peierls Centre for Theoretical Physics, Parks Road, Oxford, OX1 3PU, UK}

	\begin{abstract}
    We investigate the full doping and strain-dependent phase diagram of the normal state of magic-angle twisted bilayer graphene (TBG). Using comprehensive Hartree-Fock calculations, we show that at temperatures where superconductivity is absent
     the global phase structure can be understood based on the competition and coexistence between three types of intertwined orders: a fully symmetric phase, spatially uniform flavor-symmetry-breaking states, and an incommensurate Kekul\'e spiral (IKS) order. For small strain, the IKS phase, recently proposed as a candidate order at all non-zero integer fillings of the moir\'e unit cell, is found to be ubiquitous for non-integer doping as well. We demonstrate that the corresponding electronic compressibility and Fermi surface structure are consistent with the `cascade' physics and Landau fans observed experimentally. This suggests a unified picture of the phase diagram of TBG in terms of IKS order.
	\end{abstract}
	
 	\maketitle
 	
 	\textit{Introduction}.---
 	When two layers of graphene are stacked with a relative twist close to the `magic angle' ($\sim1^\circ$) the resulting moir\'e pattern leads to a band structure with very flat bands, enhancing correlation effects~\cite{Bistritzer2011}.
 	As the number of electrons per moir\'e unit cell  (i.e.~the filling $\nu$, measured relative to neutrality) is varied, twisted bilayer graphene (TBG) exhibits a rich array of insulating, metallic, semi-metallic, topological, and superconducting behaviour~\cite{Lu2019,Cao2018,Cao2018b,Yankowitz_2019,Sharpe_2019,Serlin900}. Much theoretical effort has been expended in attempts to formulate a consistent framework  explaining these  phenomena~\cite{po2018,xie2020weakfield,XieSub,Bultinck_2020,liu2021theories,2020CeaGuinea,Zhang2020HF,ochi2018,KangVafekPRL,Kang2020,vafek2020RG,Liu2021nematic,dodaro2018,TBG4,TBG5,TBG6,SoejimaDMRG,kwan2021kekule,PotaszMacDonaldED,zhang2021momentum,klebl2021,shavit2021theory,wu2018phonon,lian2019phonon,wu2019phononlinear,lewandowski2021,Bultinck2019mechanism,hejazi2021,parker2020straininduced,thomson2021,Christos_2020,khalaf2021charged,Chatterjee2020,cea2021electrostatic}. Three main classes of experimental data (detailed below)  guide such a framework: (zero-field) transport measurements, compressibility measurements (`cascades'), and  Landau fan diagrams. 
 	In this paper we use microscopic Hartree-Fock (HF) calculations, which crucially incorporate 
 	strain and allow for spatially modulated order, 
 	to establish a global normal state phase diagram of TBG that is consistent with all these experimental results.

 	Transport measurements above $T_c$ reveal a semimetal \cite{Park_2021,Cao2018,Cao2018b,Yankowitz_2019,Cao_2021,liu2021tuning,Zondiner_2020,uri2020mapping,saito2020independent,Das_2021,saito2021isospin,rozen2021entropic} or insulator \cite{Sharpe_2019,Serlin900,Lu2019,Stepanov_2020,Wu_2021,pierce2021unconventional,polshyn2019large,stepanov2020competing} at $\nu=0$, a metal or weak insulator at $\nu=\pm1$, correlated insulators at $\nu=\pm2$~\cite{Park_2021,Cao2018,Yankowitz_2019,Cao_2021,liu2021tuning,Sharpe_2019,Serlin900,Lu2019,Stepanov_2020,Wu_2021,Zondiner_2020,pierce2021unconventional,polshyn2019large,uri2020mapping,saito2020independent,Das_2021,saito2021isospin,rozen2021entropic,stepanov2020competing} and $\pm3$~\cite{Yankowitz_2019,Sharpe_2019,Serlin900,Lu2019,liu2021tuning,Stepanov_2020,Wu_2021,pierce2021unconventional,polshyn2019large,uri2020mapping,saito2020independent,saito2021isospin,stepanov2020competing}, and metallic behaviour~\cite{Park_2021,Cao2018,Cao2018b,Yankowitz_2019,Cao_2021,liu2021tuning,Sharpe_2019,Serlin900,Lu2019,Stepanov_2020,Wu_2021,Zondiner_2020,pierce2021unconventional,polshyn2019large,uri2020mapping,saito2020independent,Das_2021,saito2021isospin,rozen2021entropic,stepanov2020competing} at all non-integer fillings. Strong-coupling calculations predict an insulator  at $\nu=0$ \cite{Bultinck_2020,TBG4}; thus the observed semimetallic behaviour indicates that experimental TBG samples lie outside the strong coupling regime. Theoretically, it has been demonstrated that even small amounts of strain suppress strong coupling insulators at $\nu=0$~\cite{parker2020straininduced} in favour of semimetallic behaviour. Since finite heterostrain of strength $\epsilon=0.1-0.7\%$ has been measured via scanning probes~\cite{Kerelsky2019,Choi2019,Xie2019stm,Wong_2020}  in many TBG samples and is likely ubiquitous, its inclusion provides a compelling explanation of experiments at charge neutrality.

 	  	 \begin{figure*}
    \includegraphics[ width=\linewidth]{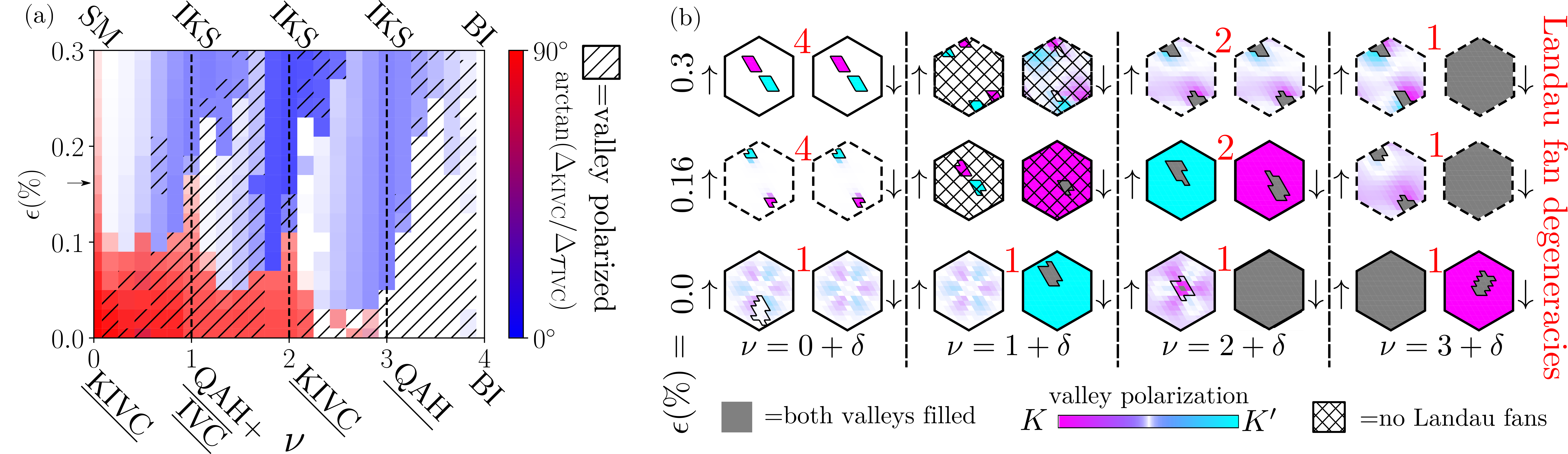}
    \caption{(a) Hartree-Fock phase diagram of TBG in filling-strain plane. % of the IVC order parameter. % of the lowest-energy HF solution. 
    Colormap intensity shows the magnitude of the IVC order parameter (white regions have no IVC). Color encodes  type of IVC via the angle $\arctan(\Delta_\textrm{KIVC}/\Delta_{\mathcal T\textrm{IVC}})$ on the `Bloch sphere', KIVC at $\vec{q}=0$ (red) and $\mathcal T\textrm{IVC}$ at non-zero $\vec{q}$ (blue). Hatching denotes valley polarization. $\mathcal T$-breaking phases are underlined. HF results are for $12\times 12$ systems, minimizing over all $\mathbf{q}$. (b) Electron-doped FSs (black lines) of both spin species ($\uparrow,\downarrow)$ near commensurate filling at three representative strains.   In IVC phases, valleys hybridize as indicated by the modulated valley polarization, yielding two split bands of which only the lower is shown. {Finite-$\mathbf{q}$  IVC order is flagged by dashed mBZ boundaries ($\mathbf{q}$ may vary).} Experimentally-measured Landau fans indicate $4,2,1$ equal-area FSs on electron-doping $\nu=0,2,3$. {HF results at $\epsilon=0.16\%$ and $\epsilon=0.3\%$ are consistent with this, but  not at $\epsilon=0$, suggesting strain is ubiquitous  in real samples. {Metallic parent states at $\nu=1$, $\epsilon=0.16\%, 0.3\%$ (hatched) do not give rise to clear Landau fans, consistent with experiments. } } 
    }
    \label{fig:main_fig}
    \end{figure*}

    \begin{table*}
\centering
\newcommand{\colskip}{\hskip 0.05in}
\renewcommand{\arraystretch}{1.15}
\begin{tabular}{
l @{\hskip 0.3in} 
c @{\colskip} 
c @{\colskip} 
c @{\colskip} 
c @{\colskip} 
c @{\colskip} 
c @{\colskip} 
c @{\colskip} 
c @{\colskip} 
c }\toprule[1.3pt]\addlinespace[0.3em]
Phase & ref. $\nu$ &
spin pol. & 
valley pol. & 
$U(1)_V$ &
$\hat{\mathcal{T}}=\tau_x\hat{\mathcal{K}}$ & 
$\hat{\mathcal{T}}'=\tau_y\hat{\mathcal{K}}$  & $P=\langle \hat c^\dagger_{\mathbf{k},\tau s\sigma }\hat c^{\phantom{\dagger}}_{\mathbf{k}',\tau's'\sigma'}\rangle$
\\ \midrule
IKS& -3&  * & 0 & \xmark & \cmark & \xmark &  $\frac{1}{8}(1+s_z)(1+\mathbf{n}_\mathbf{k}\cdot\boldsymbol{\gamma})(\delta_{\mathbf{k},\mathbf{k'}}+\mathbf{m}_\mathbf{k}^\perp\cdot{\boldsymbol{\eta}}^\perp\delta_{\mathbf{k},\mathbf{k'+q}}+m_\mathbf{k}^z\eta^z\delta_{\mathbf{k},\mathbf{k'}})$  \\
VP&-2 & 2 & 2 & \cmark & \xmark & \xmark & $\frac{1}{4}(1+s_z)(1+\tau_{z})\sigma_0\delta_{\mathbf{k},\mathbf{k'}}$\\
SH&-2 &  0 & 2 & \cmark &  \xmark & \xmark &  $\frac{1}{4}(1+\tau_{z})(1+\sigma_{z}s_z)\delta_{\mathbf{k},\mathbf{k'}}$\\
VH&-2 &  * & 0 & \cmark &  \cmark & \cmark & \ $\frac{1}{4}\tau_0(1+s_z)(1+\sigma_{z})\delta_{\mathbf{k},\mathbf{k'}}$\\
QAH&-3 &  1 & 1 & \cmark &  \xmark & \xmark & \ $\frac{1}{8}(1+s_z)(1+\sigma_z)(1+\tau_{z})\delta_{\mathbf{k},\mathbf{k'}}$\\
KIVC&-2 &  * & 0 & \xmark & \xmark & \cmark & $\frac{1}{4}(1+s_z)(1+(\cos\theta_\textrm{IVC}\tau_{x}+\sin\theta_\textrm{IVC}\tau_{y})\sigma_{y})\delta_{\mathbf{k},\mathbf{k'}}$  \\  
\bottomrule[1.3pt]
\end{tabular}
\caption{
Order parameters and representative projectors $P$ in the Chern basis at a given reference $\nu$. Doped versions of these phases appear at non-integer fillings.
Asterisks denote degenerate manifolds of states with different spin polarizations obtained by performing valley-dependent spin rotations allowed by $SU(2)_K\times SU(2)_{K'}$ symmetry. $\boldsymbol{\gamma}=(\sigma_x,\tau_z\sigma_y,\tau_z\sigma_z)$, $\boldsymbol{\eta}=(\tau_x\sigma_x,\tau_y\sigma_x,\tau_z)$ and $\mathbf n_\mathbf{k}$, $\mathbf m_\mathbf{k}$ are three-vectors ($\mathbf n_\mathbf{k}$ lies in the $x-y$ plane).  [IKS: incommensurate Kekul\'e spiral, QAH: quantized anomalous Hall state, KIVC: Kramers intervalley coherent state, VP: valley polarized state, VH/SH: valley/spin Hall states.]} 
\label{tab:phases}
\end{table*}

 	Viewing strain as a key ingredient at $\nu=0$ profoundly impacts the understanding of TBG at other commensurate fillings. In recent work we and others~\cite{kwan2021kekule} argued that a new state,  dubbed the incommensurate Kekul\'e spiral (IKS), emerges at all non-zero integer fillings in the range of modest  
 	strain invoked to explain the $\nu=0$ semimetal. Via HF analysis, we identified IKS order as an energetically favoured and experimentally consistent candidate for the metallic states at $\nu=\pm1$ and the gapped insulators  at $\nu=\pm2,\pm3$. Here, we  explore the relationship between the physics at non-integer $\nu$ and the previously-identified states at integer $\nu$.
 	
    	A stringent test for theory is to reproduce the phenomenology of the  Fermi surfaces (FSs) that emerge on doping away from  commensuration.  One such feature is   `cascade physics': the sequence of density-tuned transitions which repeats roughly each time $\nu$ is increased by one. Scanning tunneling experiments~\cite{Wong_2020,choi2021correlation,choi2021interactiondriven} show distinct changes in the excitation spectrum at each integer filling, whereas compressibility measurements show a characteristic repeating sawtooth pattern in the chemical potential $\mu$ \cite{Park_2021,Tomarken2019,Zondiner_2020,pierce2021unconventional,saito2021isospin,rozen2021entropic,yu2021correlated}: As $\nu$ approaches each positive integer from below, %the chemical potential
    	$\mu$ increases sharply, before gradually decreasing towards the next integer $\nu$. A complementary perspective is given by the `Landau fans' of field-dependent densities where  the longitudinal magnetoresistivity $\rho_{xx}(B)$  reaches a minimum, corresponding to filled Landau levels. These reveal  the number of degenerate FSs that emerge from the commensurate fillings. On the electron-doped side of {$\nu=0,2,3$, experiments find $4,2,1$ FSs} respectively~\cite{Sharpe_2019,Serlin900,Lu2019,Cao2018,Cao2018b,Yankowitz_2019,Park_2021,Stepanov_2020,Wu_2021,Zondiner_2020,uri2020mapping,saito2020independent,Saito_2021Hofstadter,saito2021isospin}. Hole-doping negative integers yields analogous scenarios 
    	due to approximate particle-hole symmetry\cite{Supplement}.
    	
    	The flavour degeneracy of TBG is crucial to understanding  cascade physics. Focusing on the two flat central bands, there are four flavours of electrons (two spins and two valleys) so that 8 electrons per moir\'e unit cell are necessary to completely fill the TBG flat bands. Empirically, flavour-symmetry breaking transitions  occur near van Hove singularities between  integer fillings \cite{xie2020weakfield}, with the density of electrons in partially filled bands resetting to zero at each integer; however, the  mechanism behind the cascades remains controversial. Although both weak- and strong-coupling approaches invoke the competition between  
    	Coulomb 
    	exchange (which favours flavour polarization) 
    	and  kinetic energy (minimized by 
    	equal flavour populations), they differ in  details. Weak-coupling theories build on the bare dispersion~\cite{Zondiner_2020} and its linearity near Dirac points; since each flavour has two Dirac cones near $K_M, K'_M$ in the moir\'e Brillouin zone (mBZ),  in this `Dirac cascade' picture the sequence of electron-doped FSs is 8, 6, 4, 2, inconsistent with measured Landau fans unless, e.g., $C_3$ symmetry is broken~\cite{Zhang2019_Landau}. In contrast, strong-coupling treatments rely on the  renormalization of the bare band structure by interactions: within HF  the Dirac cones are replaced by a large correlation dip, i.e. a minimum in the dispersion, near $\Gamma_M$~\cite{Guinea2018,rademaker2019smoothening,cea2019pinning,goodwin2020hartree,kang2021cascades,pierce2021unconventional}. On doping, this yields a single FS per flavour and hence gives  Landau fans consistent with experiment~\cite{kang2021cascades}. 
    	
Despite such successes, a strong-coupling calculation that ignores corrections from the single-particle dispersion fails other tests. 
Notably, it ignores strain, which is essential to reproduce the $\nu=0$ semimetal as well as the metallic behaviour at  $\nu=\pm 1$ seen by many experiments. 
Even for zero strain, in a strong-coupling calculation it is necessary to include a small amount of kinetic energy in order to pick out the Kramers intervalley coherent (KIVC) insulator (known to be the unstrained HF ground state at $\nu = 0,\pm 2$ for realistic interactions \cite{Bultinck_2020,TBG4}) from a degenerate manifold of flavor-symmetry-broken states. In the presence of an experimentally realistic amount of strain, the kinetic energy is further enhanced, hence requiring an intermediate-coupling analysis, as used in Ref.~\cite{kwan2021kekule} to argue that IKS order consistently explains metallic and insulating behavior at different integer $\nu$.
A natural question is whether this approach  also reproduces  the cascade physics and  Landau fan degeneracies. % that are the other experimental inputs to theory.

 	 With this motivation, 
 	 we extend the study of TBG with strain \cite{kwan2021kekule,parker2020straininduced} to the gapless states at non-integer fillings of the central bands. We demonstrate numerically that finite-strain IKS order remains stable when the system is doped away from integer filling, and analyze the Fermi surfaces and the chemical potential variation over the full range of experimentally relevant fillings and strains. We find that the electronic compressibility matches  experiment reasonably well  for {\it all} strains, even though the phase structure changes significantly for $\epsilon\gtrsim0.2\%$, suggesting that cascade physics places far  weaker constraints on  theory than previously assumed. We argue therefore that Landau fan degeneracy and the absence/presence of insulating states at integer fillings are more informative diagnostics of the underlying physics. Strain and the resulting IKS order are vital for the HF calculation to agree with both experimental probes. The ubiquity of IKS order at almost all fillings for the relevant strains suggests this is the universal normal state of TBG from which superconductivity emerges at low temperatures.

\textit{Results}.---
We perform self-consistent HF calculations of the interacting Bistritzer-Macdonald (BM) model~\cite{Bistritzer2011} projected to the central bands and without substrate potential
or non-local tunneling. We use  hopping parameters $w_{AA}=82.5$\,meV and $w_{AB}=110$\,meV and work at a twist angle of $\theta=1.1^\circ$. We include heterostrain of strength $\epsilon$ and  axis  along $\hat{\mathbf{x}}$ using the prescription of Refs.~\cite{Bi2019,parker2020straininduced,kwan2021kekule}. We verified that variations in twist angle and  strain axis do not qualitatively change the phase diagram. The Hamiltonian has approximate particle-hole symmetry, allowing us to restrict discussion to positive $\nu$~\cite{Supplement}. We use the dual-gate screened Coulomb interaction $V(q)=\frac{e^2}{2\epsilon_0\epsilon_r q}\tanh{qd}$ with screening length $d=25\,\textrm{nm}$ and relative permittivity $\epsilon_r=10$. To avoid double-counting interactions, we subtract off a density matrix corresponding to
decoupled graphene layers at charge neutrality~\cite{XieSub,Bultinck_2020}. We show that other subtraction schemes lead to similar results (See Supplement\cite{Supplement}). Further HF details are in Ref.~\cite{kwan2021kekule}.

Diagonalizing the sublattice operator (ignoring the weak dispersion) defines 
the `Chern basis' $\hat{c}^\dagger_{\mathbf{k},\tau s\sigma}$ for the  eight central bands~\cite{Bultinck_2020},  which  can then be labelled by valley, spin, and sublattice polarization (with Pauli matrices $\tau_\mu$, $s_\mu$ and $\sigma_\mu$ respectively),
with the Chern number of each band  given by $C=\sigma_z\tau_z$. In the strong coupling limit, the integer-filling ground states 
are uniform ferromagnets  in this basis, in analogy with quantum Hall ferromagnetism (QHFM)~\cite{Bultinck_2020,TBG4}. However, on including  finite strain  the system  leaves the strong coupling limit, and hosts new phases: A fully-symmetric phase and the IKS order. 
All these orders can be captured by the one-particle density matrix

\begin{equation}
\langle \hat c^\dagger_{\mathbf{k}-\tau\mathbf{q}/2,\tau s\sigma }\hat c^{\phantom{\dagger}}_{\mathbf{k}-\tau'\mathbf{q}/2,\tau's'\sigma'}\rangle=P_{\tau s\sigma;\tau's'\sigma'}(\mathbf{k}),
\end{equation}
with $\text{Tr}\,P=(\nu+4)N_1N_2$, where $N=N_1N_2$ is the number of moir\'e unit cells. We have shifted the momenta such that we always hybridize electrons with momentum $\mathbf{k}-\tau\mathbf{q}/2$ in valley $\tau$ with electrons with momentum $\mathbf{k}-\tau'\mathbf{q}/2$ in valley $\tau'$. That way intravalley ($\tau=\tau'$) hybridization occurs at equal momenta while intervalley ($\tau\neq\tau'$) coherence (IVC) occurs with relative momentum $\mathbf{q}$. We
choose the HF solution with the lowest energy for any $\mathbf{q}$ in the 
mBZ. 
We find two types of IVC states: Kramers-IVC (KIVC) order \cite{Bultinck_2020} at $\mathbf{q}=0$ and time-reversal symmetric $\mathcal{T}$IVC order at variable $\mathbf{q}\neq 0$ (this is the so-called IKS order \cite{kwan2021kekule}). We can define the spinless time-reversal symmetry $\mathcal T=\tau_x\mathcal{K}$ and the related antiunitary symmetry $\mathcal T'=\tau_y\mathcal{K}$, where $\mathcal{K}$ denotes complex conjugation. The KIVC order parameter $\Delta_\textrm{KIVC}=\tau_{x,y}\sigma_y$ then satisfies $\mathcal T'$ whereas $\mathcal{T}$IVC with $\Delta_{\mathcal T\textrm{IVC}}=\tau_{x,y}\sigma_x$ respects $\mathcal T$.

Fig.~\ref{fig:main_fig}(a) shows a global phase diagram of TBG in the filling-strain plane. We find the different types of order listed in Tab.~\ref{tab:phases}. Between $\nu=0$ and $\nu=2$ we find KIVC order at $\mathbf{q}=0$ for small strains (red regions). As noted, this is known to be the zero-strain HF ground state at $\nu=0$ and $\nu=2$. At $\nu=0$ the ground state switches from the insulating KIVC to a semimetallic state (SM) at a critical value of strain (larger than $\epsilon=0.3\%$ for the chosen parameters), as previously reported  \cite{parker2020straininduced}. This semimetal 
persists to finite doping as a symmetric (S) metal (white region). For modest values of strain and mostly on the hole-doped side of the integers, we find IKS order at finite  $\vec q$ (blue regions) that was previously reported only at integer fillings \cite{kwan2021kekule}. Finally, there are other generalized ferromagnets besides the KIVC: quantized anomalous Hall (QAH), valley polarized (VP), valley Hall (VH), and spin Hall (SH) \footnote{We term this a spin Hall state since opposite spins have opposite Chern number, leading to a spin Hall current. However this state is \textit{not} a topological insulator and does not have spin-Kramers time reversal symmetry.} states (white regions) that are in close energetic competition with the IKS solution. 
Between the integer fillings, we find first-order phase transitions where the ground state crosses over between an IKS solution and a VP solution (between $\nu=1-2$ and $\nu=3-4$) or a VH/SH solution (between $\nu=2-3$; the two solutions are degenerate in our calculation, however intervalley Hund's coupling favours the VH state \cite{Supplement}). VP can also coexist with the KIVC and IKS away from the integer fillings. {$\mathcal{T}$-breaking is ubiquitous at zero strain, but is almost completely absent 
for $\epsilon=0.3\%$.}

We have verified that the order parameters remain nonzero in finite-temperature HF for temperatures up to  $\sim 50~K$  \cite{Supplement}, although those that break $U(1)$ symmetries  will only have algebraic correlations once fluctuations beyond mean-field are included. 
Consequently, the phases studied here are possible parent states for the superconductors that emerge in experiments below  $T_c\lesssim 5$K \cite{Lu2019,Cao2018,Cao2018b,Yankowitz_2019}. 

Depending on the strain, different pictures emerge for the cascades. 
{At zero strain, at $\nu=0$ the system begins in KIVC, with IVC in both spin species. Upon doping,  down spins continue to have IVC order, while  up spins  valley polarize. At $\nu=1$,  down spins are in a KIVC insulator while up spins form a QAH state. We move to $\nu=2$ by doping  up spins until they are fully filled, while down spins are in KIVC. Doping beyond $\nu=2$,  KIVC order is destroyed in the down spins, which now become valley polarized and eventually form QAH state at $\nu=3$.}

At finite strain,  doping the semimetal at neutrality gives four FSs from the two spins and two valleys. The states are filled from the Hartree minima, which shift away from $\Gamma_M$  by equal and opposite amounts in each valley, consistent with $C_2$ symmetry. At a critical filling, the Fermi seas of the two valleys start overlapping, signaling the onset of finite-$\mathbf{q}$ IVC order (IKS). By the time the next integer filling is reached, the IKS order has proliferated throughout the mBZ (except near the Hartree minima, where  electrons remain valley-polarized). Therefore we can view the emergence of IKS order at critical fillings as a weak-coupling instability of a FS from interaction-renormalized bands. For $\epsilon=0.1-0.2\%$ there are regions on the electron-doped sides of %the integer fillings
$\nu=0,1,2,3$ where the ground state lacks any IVC order. {By counting the number of (equal area) FSs (Fig.~\ref{fig:main_fig}(b)) we obtain Landau fan degeneracies of ${4,2,1}$ at $\nu=0,2,3$ respectively}. For larger strains the ground state always includes some IKS order (except for fillings close to $\nu=0,4$) with wavevector $\mathbf{q}$ whose optimal value varies throughout the mBZ as the filling changes (see SM~\cite{Supplement} for $\epsilon$ and $\nu$ dependence of $\vec{q}$). For {the `Kekul\'e cascades'}, the Landau fan degeneracies at $\nu=0,2,3$ are $4,2,1$ respectively. Since we find a metal at $\nu=1$, no clear Landau fans would emanate from this filling, consistent with all experiments under normal conditions\cite{Lu2019,Cao2018b,Yankowitz_2019,uri2020mapping,saito2020independent}. A striking feature seen in experiments is the asymmetry of the Landau fans: At $\nu=\pm2,\pm3$, the Landau fans are only seen for doping away from charge neutrality\cite{Sharpe_2019,Serlin900,Lu2019,Cao2018,Cao2018b,Yankowitz_2019,Park_2021,Stepanov_2020,Wu_2021,Zondiner_2020,uri2020mapping,saito2020independent,Saito_2021Hofstadter,saito2021isospin}. By calculating the density of states at the Fermi surface, we see that for doping towards charge neutrality the bands are very flat and there is no sharp Fermi surface consistent with the absence of Landau fans. We see this phenomenology for any value of the strain (see Supplement \cite{Supplement}).

 	 \begin{figure}
    \includegraphics[ width=0.8\columnwidth]{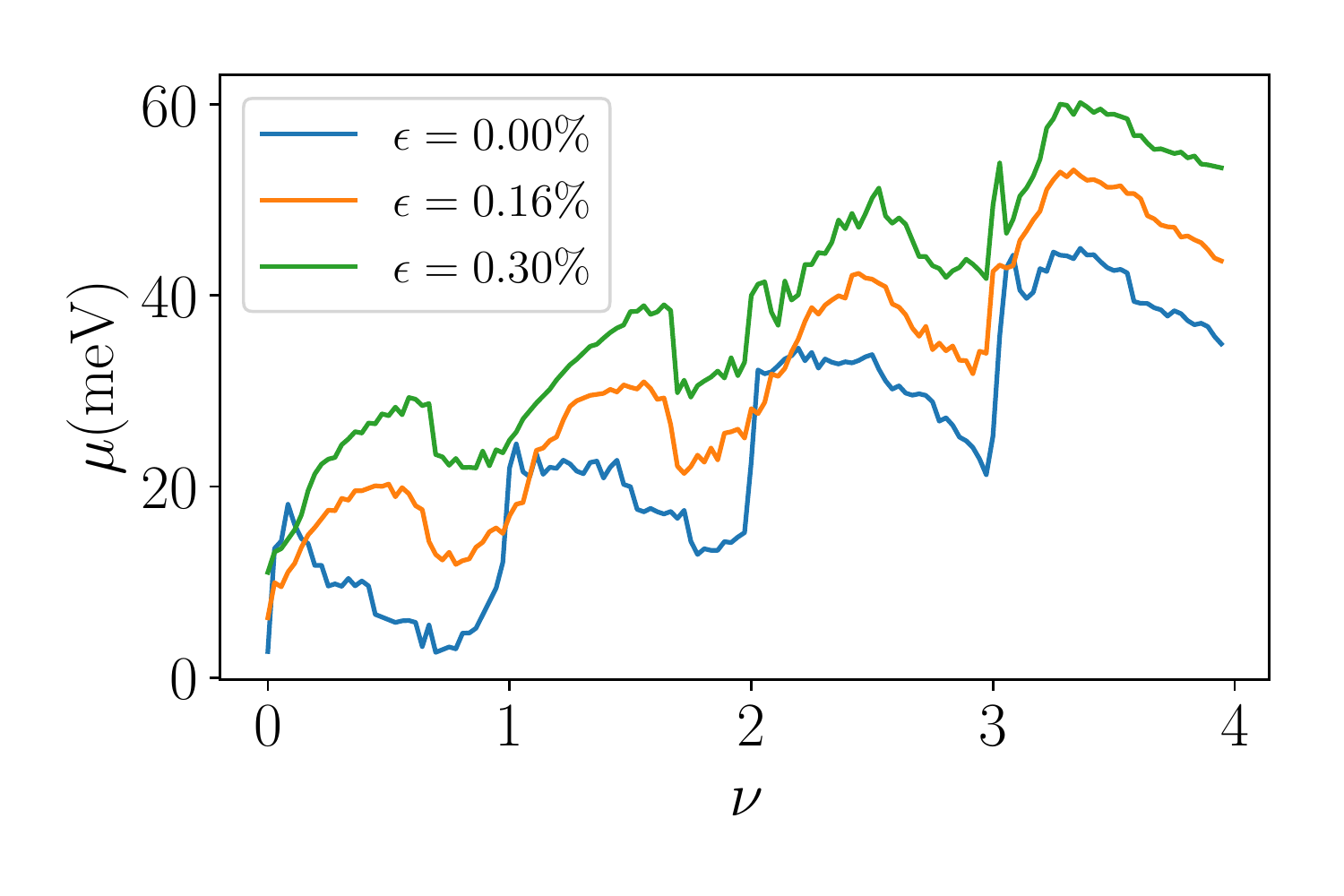}
    \caption{Chemical potential $\mu$ as a function of filling $\nu$ for three different strain values, offset by 5~meV for clarity. HF calculation for a $12\times 12$ system with steps of $\Delta\nu=1/36$.}
    \label{fig:Mu}
    \end{figure}

The traces of the chemical potential in Fig.~\ref{fig:Mu} show a sawtooth pattern that is consistent with the experimental observations reported in Refs.~\cite{Park_2021,Zondiner_2020}. The phases on the hole-doped side of integer filling are more compressible than those on the electron-doped side. For larger strains, the chemical potential is a smoother function of filling closer to the `square root' shape previously thought to be a unique signature of an underlying Dirac description~\cite{Zondiner_2020}. To understand this, note that at the largest values of the strain ($\sim 0.3\%$) there is a continuous range of IKS. The IKS state evolves smoothly, with a gradually changing $\vec q$, leading to a smooth variation in the chemical potential (up to finite-size effects). Furthermore, the chemical potential increases by 40~meV between $\nu=0$ and $\nu=4$, consistent with Refs.~\cite{Tomarken2019,Park_2021,Wong_2020,Zondiner_2020,kang2021cascades}. We note however that the chemical potential traces are relatively insensitive to the underlying phases, and the resetting of the chemical potential at integer fillings is a generic feature both of our HF studies at \textit{all} strains and also of zero-strain calculations at both  weak~\cite{Zondiner_2020} and strong~\cite{kang2021cascades} coupling.  In contrast, Landau fans are sensitive to the number and structure of the FSs and are hence better able to distinguish between competing scenarios.

\textit{Conclusions}.---
In this work we have fleshed out the full HF phase diagram of TBG above $T_c$ for any filling $\nu$ of the central bands and and as a function of strain $\epsilon$, and demonstrated that it captures key experimental features of TBG (modulo superconductivity). Reproducing the correct Landau fans and semimetallicity at $\nu=0$ requires an intermediate coupling picture with non-zero strain. A weak coupling description fails to reproduce the observed Landau fans, since the two Dirac points per  non-interacting band doubles the  number of Fermi surfaces relative to experiment. Absent  strain, a strong coupling approach can reproduce the correct Landau fans, but predicts a gapped state at $\nu=0$. 
This justifies our inclusion of both strain and realistic interactions as a necessary prerequisite to fully match  experiments. 

One of our key messages is that the normal state phase diagram of TBG can be understood in terms of three types of competing states : A symmetry-preserving metal 
and two classes of symmetry-breaking orders ---  IKS and  a set of generalized ferromagnets. Without strain, the generalized ferromagnetic states are exact ground states \cite{TBG4} at integer fillings in an idealized limit of the Hamiltonian.
 In accord with this, we  find that these states and their doped descendants describe the entire range of fillings at zero strain. However,  strain is ubiquitous in experimental samples and upon its inclusion  realistic TBG departs from the limit where generalized ferromagnets are ground states. This leads to the two types of states we find that do not lie within the manifold of generalized ferromagnets: The completely symmetric metal and the IKS state. For relatively modest strains of $\epsilon\gtrsim0.3\%$, IKS order exists for almost the entire range of fillings $\nu$. Due to its variable wavevector $\mathbf{q}$, the IKS order readily adjusts to changes in parameters, explaining its ubiquity in the phase diagram. This underscores the importance of an experimental search for this order. Furthermore, our study suggests that a doped IKS state could play the role of a parent to the superconducting order that emerges below $T_c$. The angle of the {optimal} IKS wavevector $\mathbf{q}$ varies as a function of filling~\cite{Supplement},  potentially providing an explanation for the rotating nematicity observed near $T_c$~\cite{Cao_2021}. A theoretical investigation of a superconducting mechanism from an IKS parent state is clearly warranted, and may provide the final piece of the puzzle of competing orders in TBG.

\begin{acknowledgements}
\textit{Acknowledgements}.---This research was partially supported by the NCCR MARVEL, a National Centre of Competence in Research, funded by the Swiss National Science Foundation (grant number 182892). NB is supported by a senior postdoctoral research fellowship of the Flanders Research Foundation (FWO). We acknowledge support from the European Research Council under the European Union Horizon 2020 Research and Innovation Programme, Grant Agreement No. 804213-TMCS and  from EPSRC Grant EP/S020527/1. Statement of compliance with EPSRC policy framework
on research data: This publication is theoretical work
that does not require supporting research data.
\end{acknowledgements}

\bibliography{bib}

\newpage
\clearpage

\begin{appendix}
\onecolumngrid
	\begin{center}
		\textbf{\large --- Supplementary Material ---\\ Global Phase Diagram of the Normal State of  Twisted Bilayer Graphene}\\
		\medskip
		\text{Glenn Wagner, Yves H. Kwan, Nick Bultinck, Steven H. Simon and S.A.~Parameswaran}
	\end{center}
	
		\setcounter{equation}{0}
	\setcounter{figure}{0}
	\setcounter{table}{0}
	\setcounter{page}{1}
	\makeatletter
	\renewcommand{\theequation}{S\arabic{equation}}
	\renewcommand{\thefigure}{S\arabic{figure}}
	\renewcommand{\bibnumfmt}[1]{[S#1]}

\section{Landau fans}
In this section we elaborate on the Fermi surface (FS) and Landau fan structure for moderate/large strains at very small electron dopings $\delta$ above the positive integer fillings. For sufficiently large strains, the interaction and strain renormalized electron bands (in the absence of IVC) are characterized by low energy lobes~\cite{kwan2021kekule}, one per spin and valley. Numerically we find that the HF solution at $\nu=0+\delta$ preserves all symmetries, yielding four degenerate FS from filling up the lobes. The HF solution for $\nu=3+\delta$ can be either an IKS or a generalized ferromagnet---in either case there is only one empty band to fill leading to a singly degenerate Landau fan. The situation at fillings $\nu=1+\delta$ and $2+\delta$ is more subtle. For moderate strain $\epsilon\simeq 0.1-0.2\%$, the HF ground state according to Fig.~1a is a $U(1)$-preserving generalized ferromagnet. In this case the FS degeneracy is $3$ and $2$ respectively, from doping the lobes. At yet larger strains, the state at $\nu=1+\delta$ becomes an IKS. Here, the presence of a single electron IKS band hybridizes two of the lobes in one spin sector. As a result, the only low-energy states that can be filled reside in the two lobes of the other spin sector, leading to a Landau fan degeneracy of 2, consistent with experiments under hydrostatic pressure \cite{Yankowitz_2019} (note that at $\nu=1$ the state is no longer gapless at larger strains, so the Landau fans would not emanate exactly from here). For $\nu=2+\delta$, we obtain the IKS with doubly-degenerate Landau fans, where two unfilled bands remain after formation of two electron IKS bands (one in each spin sector). 

\begin{figure}[h]
    \centering
    \includegraphics[width=\textwidth]{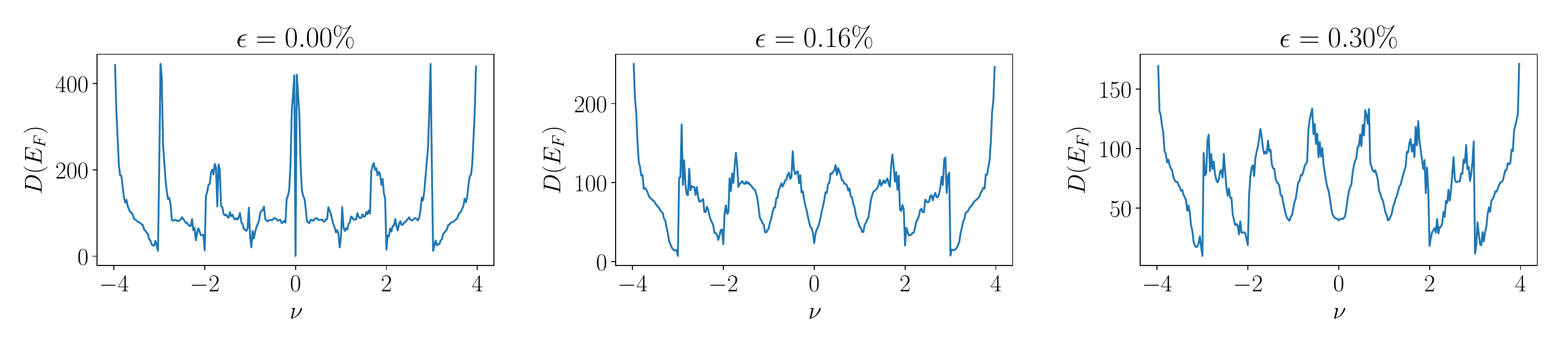}
    \caption{We show the density of states at the Fermi surface (calculated from the $12\times12$ bandstructures with a Gaussian broadening of 0.25meV) for three representative values of the strain. }
    \label{fig:DOS}
\end{figure}

A striking feature seen in experiments is the asymmetry of the Landau fans: The Landau fans at $\nu=\pm2,\pm3$ are only seen pointing away from charge neutrality. In Fig.~\ref{fig:DOS} we show the density of states at the Fermi surface which also shows a marked asymmetry at $\nu=\pm2,\pm3$: Doping away from charge neutrality we see a gradual increase in the density of states. However, doping towards charge neutrality there is a sharp jump in the density of states corresponding to very flat bands.  The same phenomenology was seen in Ref.~\cite{kang2021cascades}. The presence of these extremely flat bands means there will be no sharp Fermi surface upon doping towards charge neutrality and hence no clear Landau fans.

Another significant feature of the density of states is the behaviour around charge neutrality. At zero strain, the doped KIVC state has very flat bands and therefore we observe a sharp peak in the density of states. For finite strain, the bands that arise when doping the semimetal at charge neutrality are dispersive. This is therefore a clear spectroscopic signature of the presence or absence of the KIVC state at neutrality.

\section{Lifting of the degeneracy between VH and SH solution}
The VH state (with indefinite spin since we can use $SU(2)_K\times SU(2)_{K'}$ to independently rotate the spins in the two valleys) and the SH state (which can be obtained from the spin-unpolarized VH state by applying $C_2$ to one spin species), are exactly degenerate at this level of study. However in VH, spins in the two valleys can freely rotate to take advantage of either sign of intervalley Hund's coupling, lowering its energy relative to SH when this is included.

\section{Subtraction scheme}

\begin{figure}[h]
    \centering
    \subfloat[\centering decoupled graphene scheme]{{\includegraphics[width=6cm]{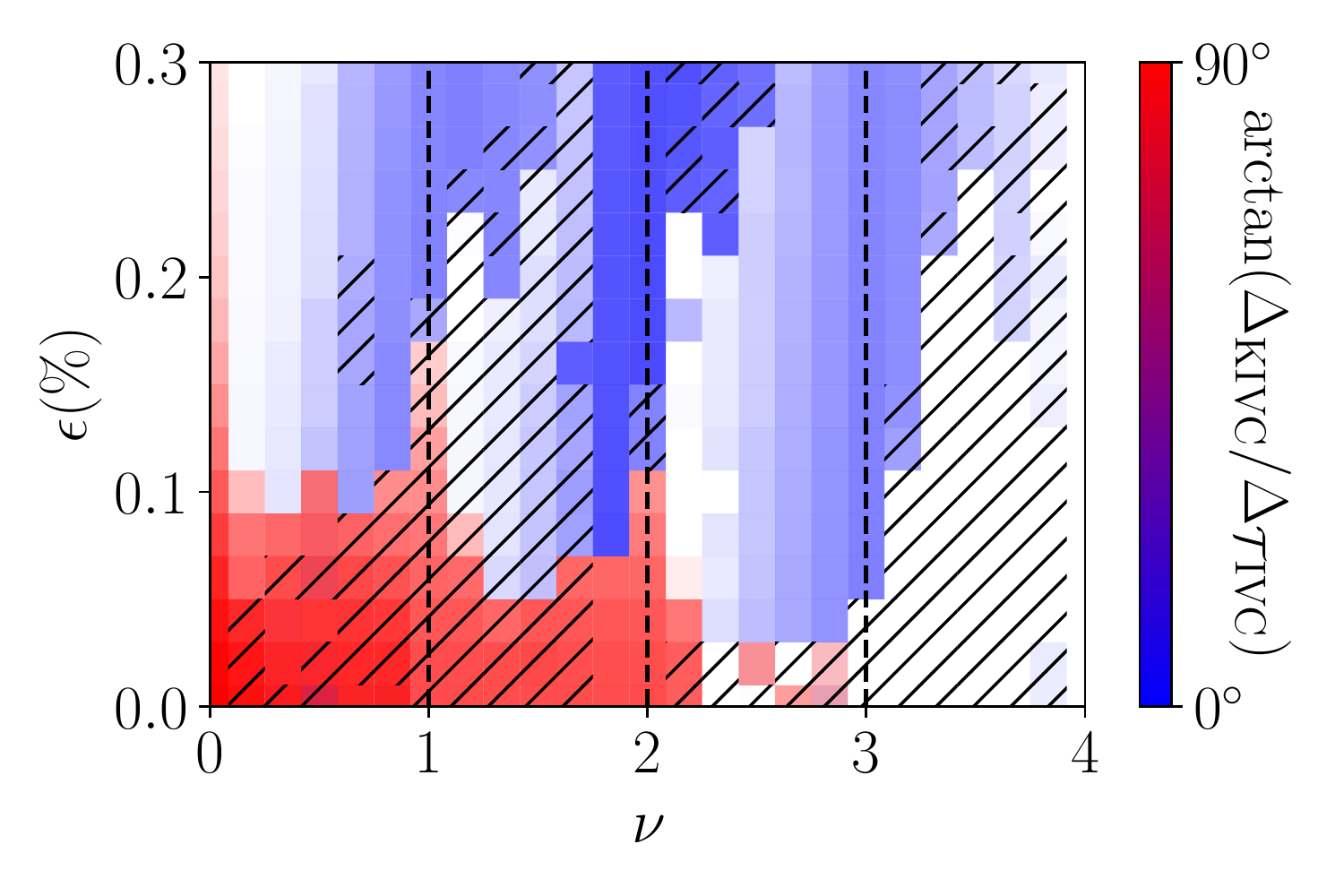} }}%
    \subfloat[\centering average scheme]{{\includegraphics[width=6cm]{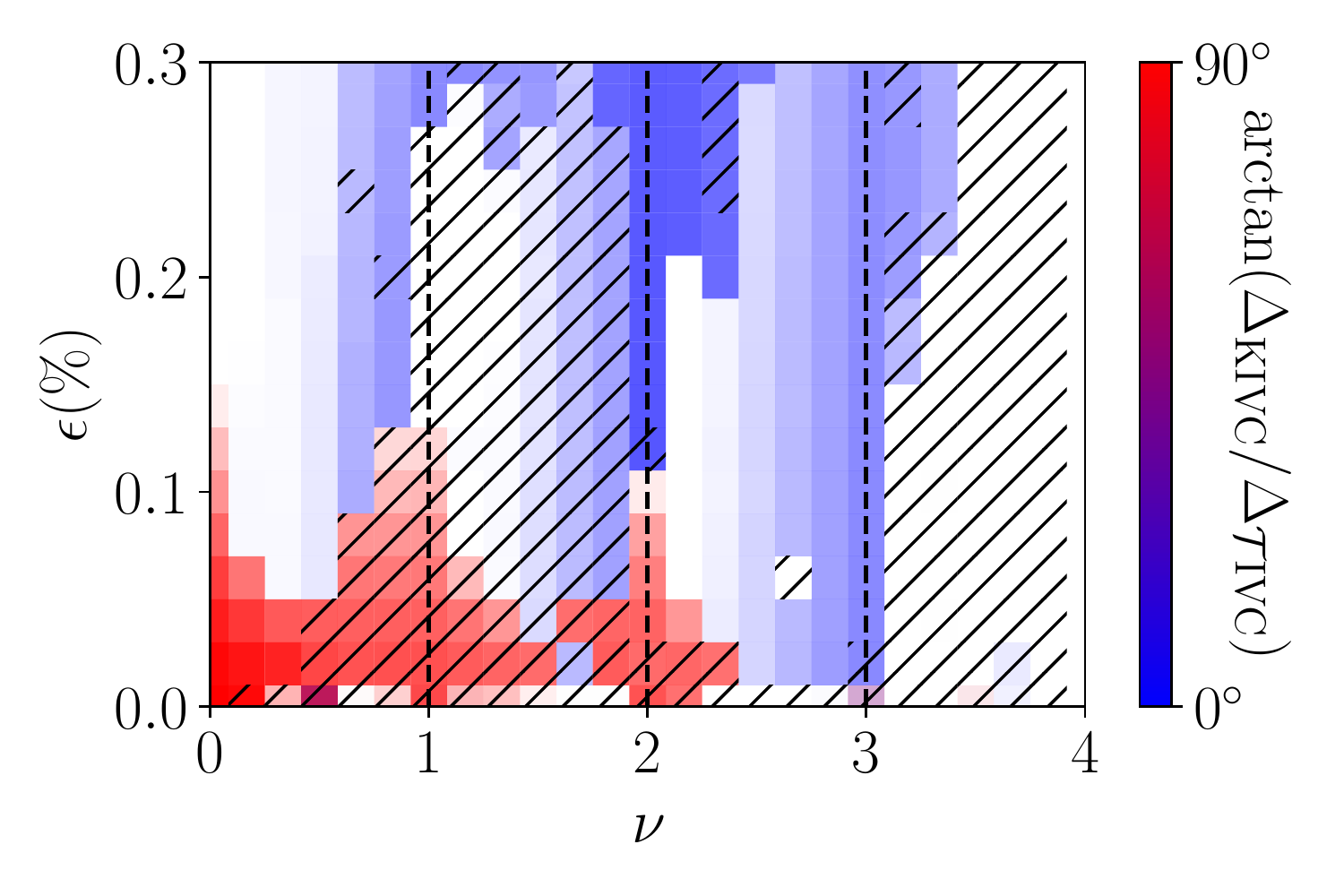} }}%
    \subfloat[\centering charge neutrality scheme]{{\includegraphics[width=6cm]{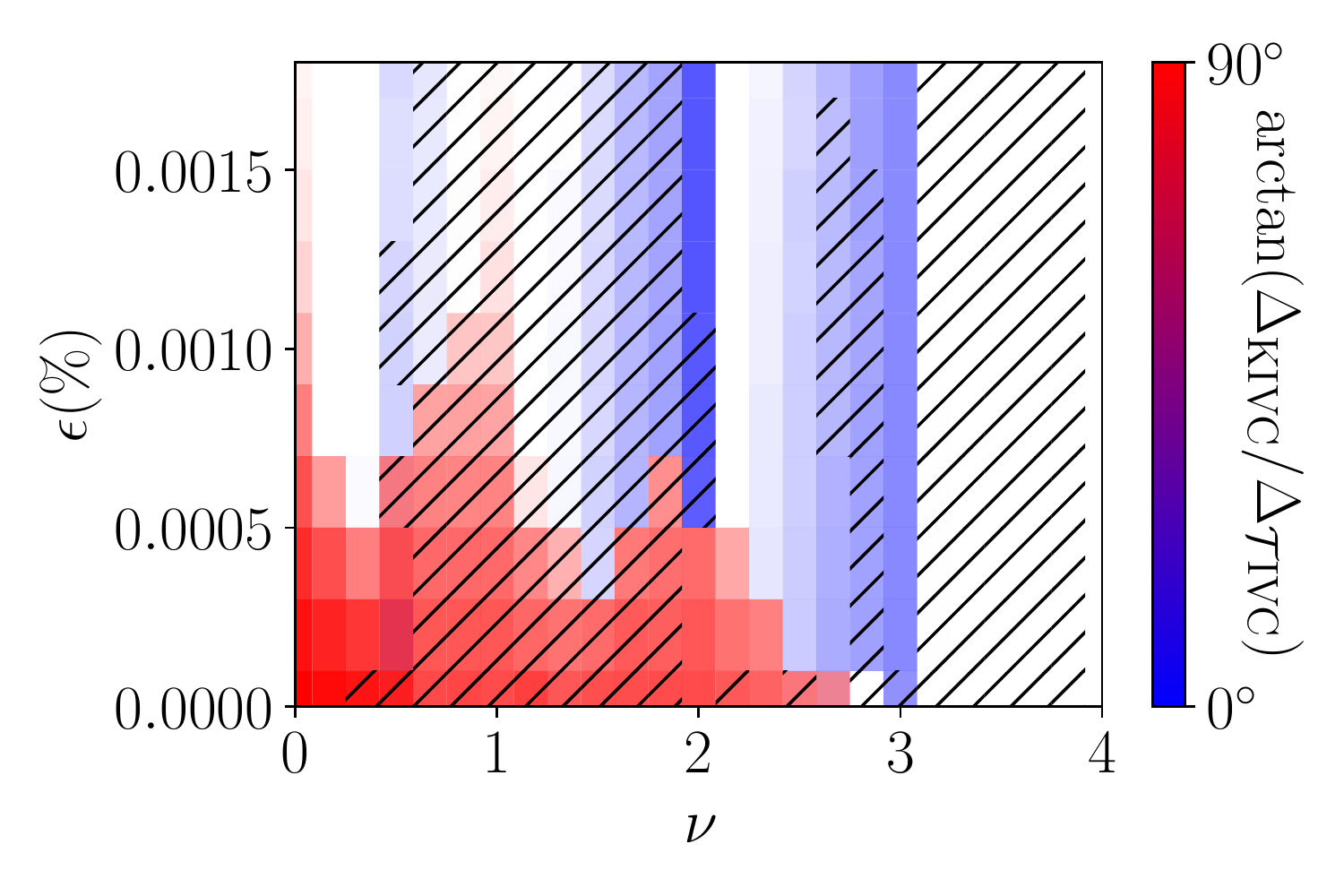} }}%
    \caption{We reproduce the phase diagram from the main text using three different schemes for taking the interactions into account. In order to avoid double counting the interactions we subtract off a reference projector. In (a) the reference projector subtracted off is the projector corresponding to two decoupled graphene layers at charge neutrality. This is the subtraction schemes used for all other calculations in the main text and supplement. In (b) the reference projector consists of half filling every state in the central band Hilbert space. In (c) we take TBG at charge neutrality as the reference projector. Note the different $y$-axis scale on this figure compared to the other schemes.}%
    \label{fig:schemes}%
\end{figure}

When performing Hartree-Fock calculations on TBG one needs to be careful not to double-count the interactions. The values of the parameters in the BM model are extracted from DFT calculations which  already include interaction contributions. Therefore we subtract a term from the HF Hamiltonian such that at a certain reference point, the contributions from the HF treatment of the interactions vanish. There are different choices of this reference projector in the literature:
\begin{enumerate}[label=(\alph*)]
    \item decoupled graphene scheme\cite{Bultinck_2020,XieSub}: two decoupled graphene layers at charge neutrality
    \item average scheme\cite{kang2021cascades,TBG3}: half filling of every state in the central band Hilbert space
    \item charge neutrality scheme\cite{hejazi2021}: TBG at charge neutrality
\end{enumerate}
For all calculations and figures in the main text and supplement we use the decoupled graphene scheme, except in Fig.~\ref{fig:schemes}, where we show the results for the two other schemes used in the literature. Fig.~\ref{fig:schemes} shows that schemes (a) and (b) yield quantitatively very similar results, while scheme (c) only yields qualitatively similar results. In particular, the values of strain at which the IKS state first appears are much smaller for scheme (c) compared to schemes (a) and (b). This observation was already made in Ref.~\cite{kwan2021kekule}, where an explanation of this phenomenology is given.

\section{Supplementary figures}

\begin{figure}[h]
    \centering
    \includegraphics[width=\textwidth]{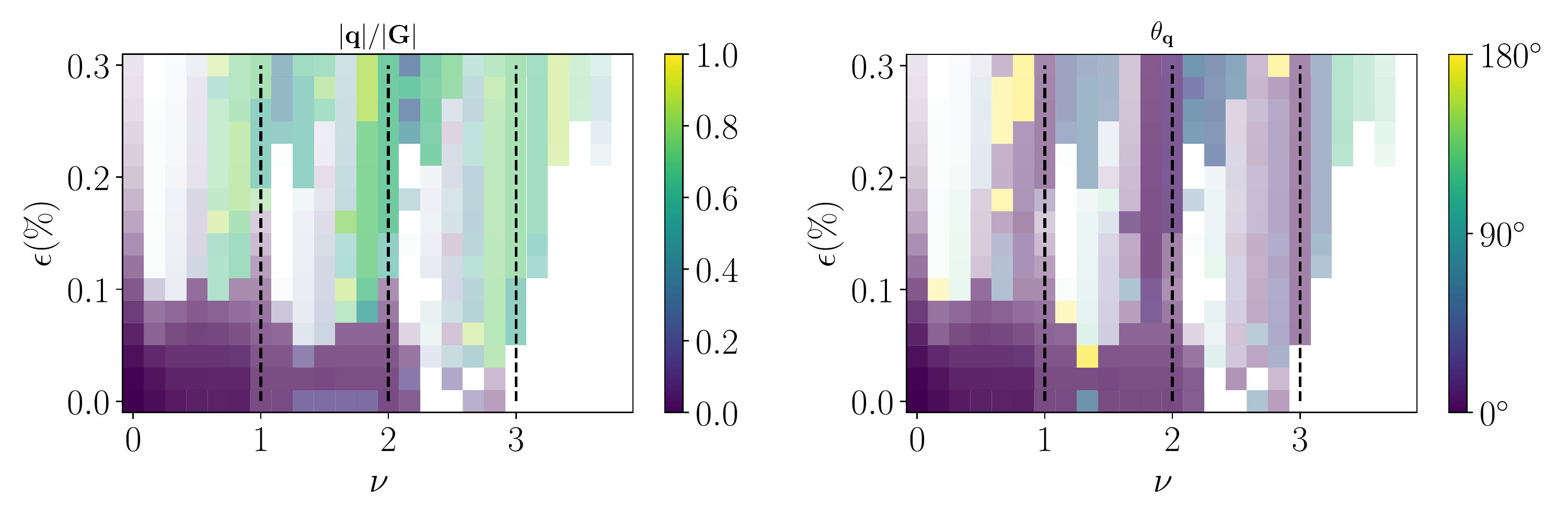}
    \caption{Magnitude and direction of the wavevector $\mathbf{q}$ of the IVC order of the lowest energy HF solution. The depth of the colour indicates the strength of the IVC order. White regions indicate regions without any IVC order, where $\mathbf{q}$ has no meaning. KIVC order (red in Fig.~\ref{fig:pols}d) is found at $\mathbf{q}=0$, whereas the $\mathcal T$IVC order (blue in Fig.~\ref{fig:pols}d is found at finite $\mathbf{q}$, i.e.~we have an IKS solution. Both the magnitude and direction of $\mathbf{q}$ evolve as a function of doping, which is consistent with the IKS dispersion having a broad minimum (as already noted in Ref.~\cite{kwan2021kekule}).}
    \label{fig:q}
\end{figure}

\begin{figure}[h]
    \centering
    \includegraphics[width=\textwidth]{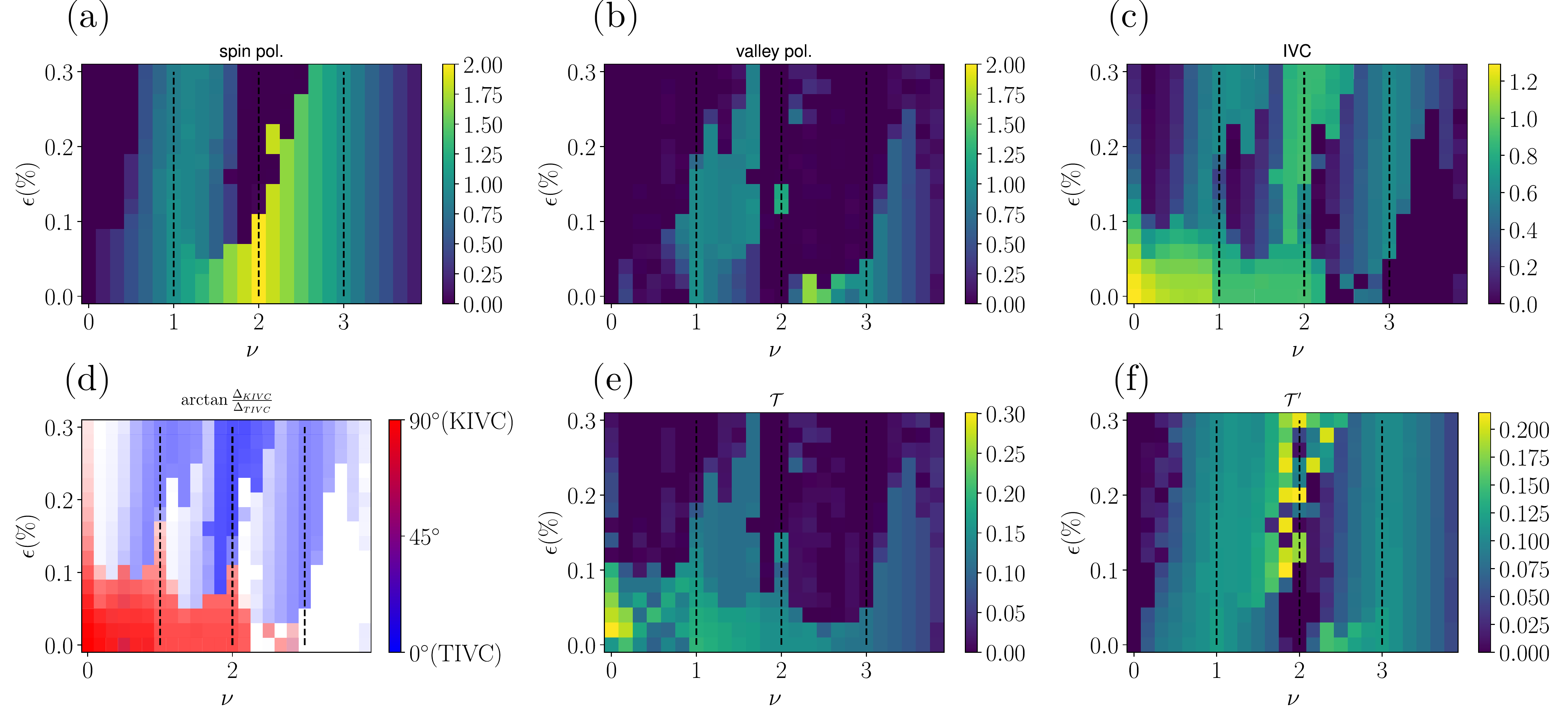}
    \caption{Plots of various order parameters for the lowest energy HF solution: (a) spin polarization,  (b) valley polarization, (c) IVC order parameter (this measures the norm of the $U_V(1)$ breaking part of the ground state density matrix), (d) IVC characterization, (e) $\mathcal{T}$ breaking and (f) $\mathcal{T}'$ breaking. There are two sources of time-reversal symmetry breaking: Valley polarization and KIVC order, as can be seen by comparing (b), (d) and (e). The IKS state preserves $\mathcal T$ even in the presence of doping, whereas the KIVC solution only preserves $\mathcal T'$ when $\nu=0,2$. }
    \label{fig:pols}
\end{figure}

\begin{figure}[h]
    \centering
    \includegraphics[width=1\textwidth]{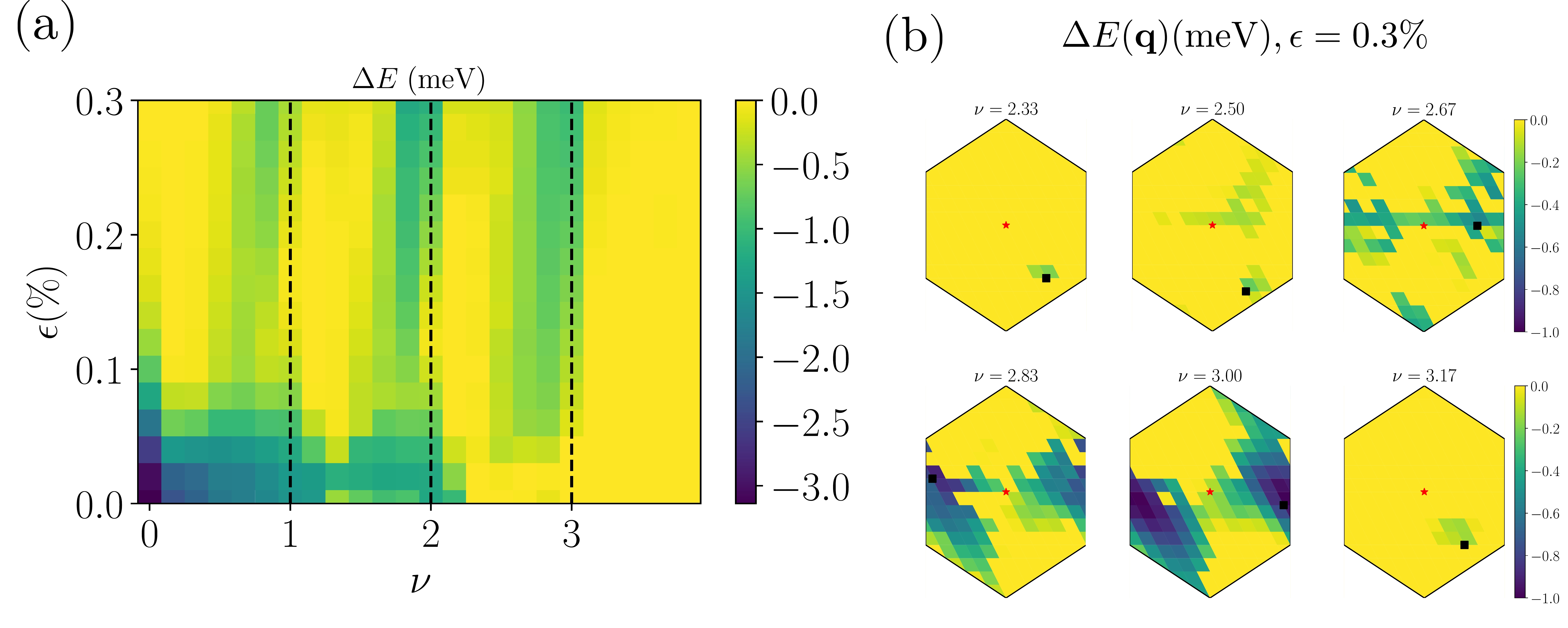}
    \caption{(a) Energy difference between the lowest energy solution and the lowest energy solution not allowing for IVC: $\Delta E=(E_\textrm{allowing for IVC}-E_\textrm{no IVC})/(N_1N_2)$. $\Delta E<0$ indicates that a solution with IVC is the lowest-energy solution. The IKS is most robust on the hole-doped side of integer $\nu$ (three green `columns'). On the electron-doped side of integer $\nu$, the IKS state is closely competing with states without any IVC (yellow regions). (b) Dispersion relation of the IKS state around $\nu=+2$ for $\varepsilon=0.3\%$ strain. The $\Gamma$ point in the Brillouin zone is denoted by a red star, the optimal IKS wavevector $\mathbf{q}_0$ is denoted by a black square.}
    \label{fig:Delta_E}
\end{figure}

\begin{figure}[h]
    \centering
    \includegraphics[width=0.5\textwidth]{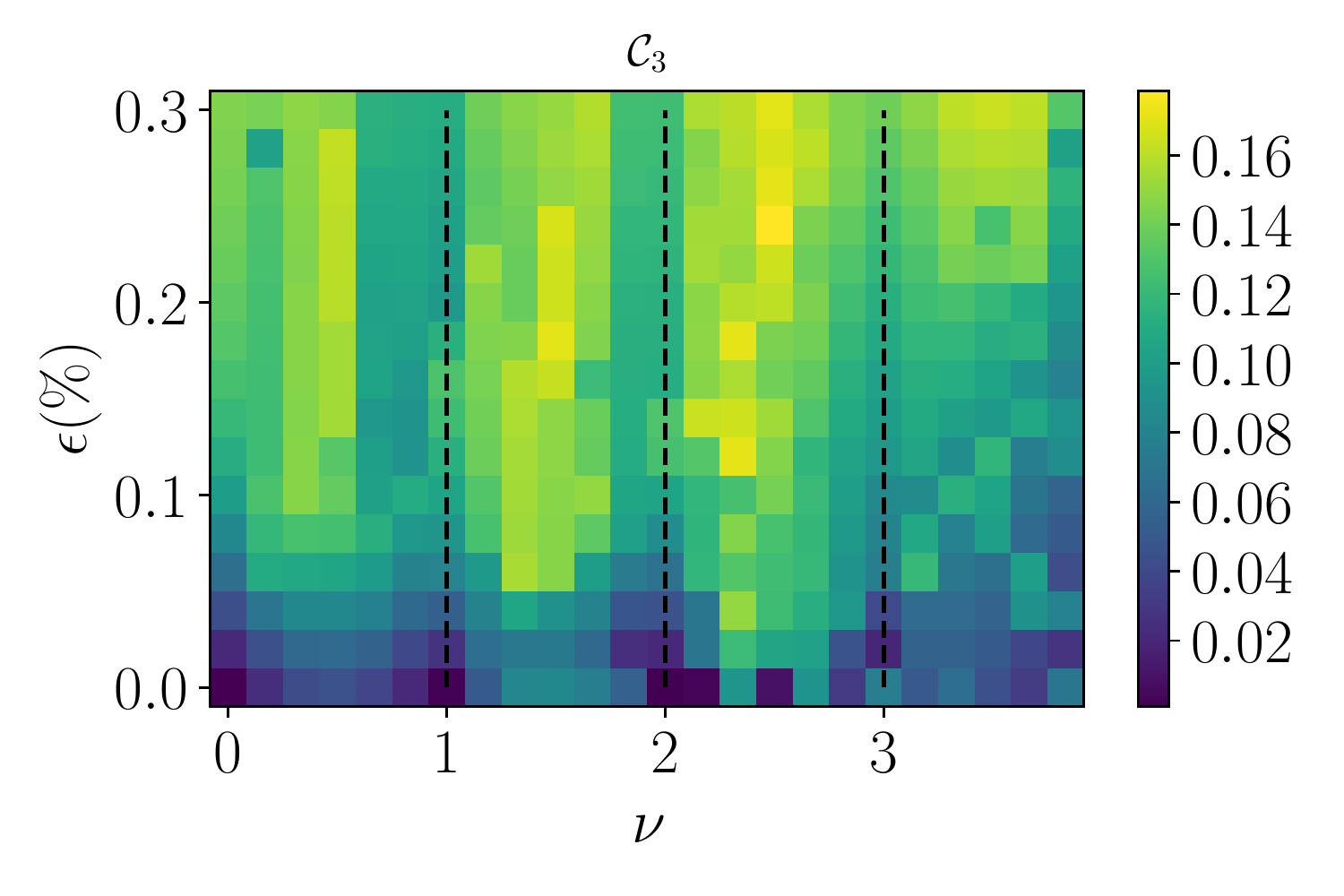}
    \caption{$\mathcal{C}_3$ breaking order parameter. We quantify the $\mathcal{C}_3$ breaking of the HF bandstructure by calculating $[\sum_{\mathbf k} |E(\mathbf{k})+\omega E(\mathcal R\mathbf{k})+\omega^2 E(\mathcal R^2\mathbf{k})|]/[3\sum_{\mathbf k} |E(\mathbf{k})|]$, where $\omega=e^{2\pi i/3}$ and $\mathcal{R}$ is a $\mathcal{C}_3$ rotation matrix. Strain breaks $\mathcal{C}_3$ explicitly and hence increases the magnitude of the $\mathcal{C}_3$ breaking. The $\mathcal{C}_3$ breaking tends to be smallest for the insulating states at integer $\nu$ and larger for the the metallic states for non-integer $\nu$.}
    \label{fig:C3}
\end{figure}

\begin{figure}[h]
    \centering
    \includegraphics[width=\textwidth]{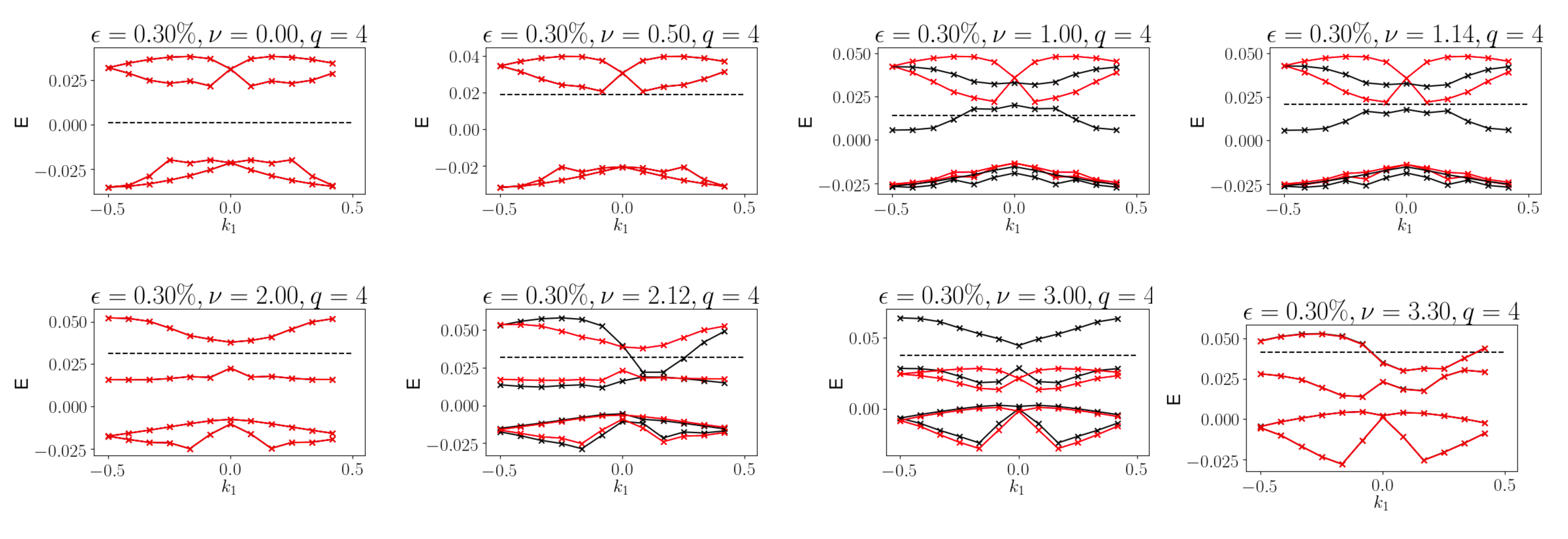}
    \caption{Band structures of the ground state for different fillings for a $12\times12$ system. Red and black bands distinguish the two spin species, the dashed line indicates the position of the chemical potential. The bandstructures demonstrate that IKS order in one spin species can coexist with a valley polarized state in the opposite spin species (as in the bandstructure at $\nu=2.12$). In these plots we fixed the IKS wavevector to be $q_1/|\mathbf{G}|=4/12=1/3$.}
    \label{fig:bs}
\end{figure}

\begin{figure}[h]
    \centering
    \includegraphics[width=\textwidth]{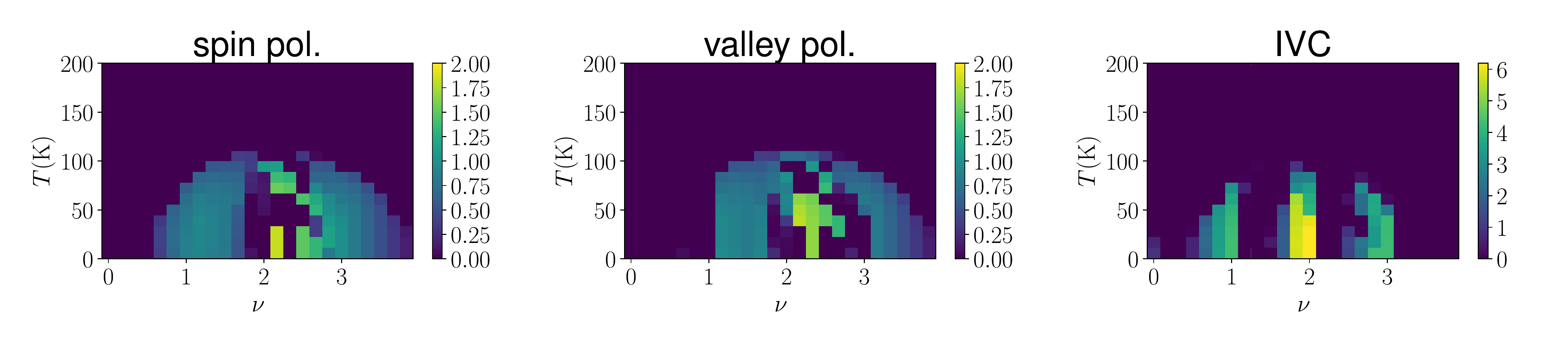}
    \caption{Order parameters from finite temperature HF at fixed strain $\epsilon=0.3\%$. The IKS state and the symmetry breaking states persist up to temperature over 50K.}
    \label{fig:finite_T}
\end{figure}

\begin{figure}[h]
    \centering
    \includegraphics[width=0.8\textwidth]{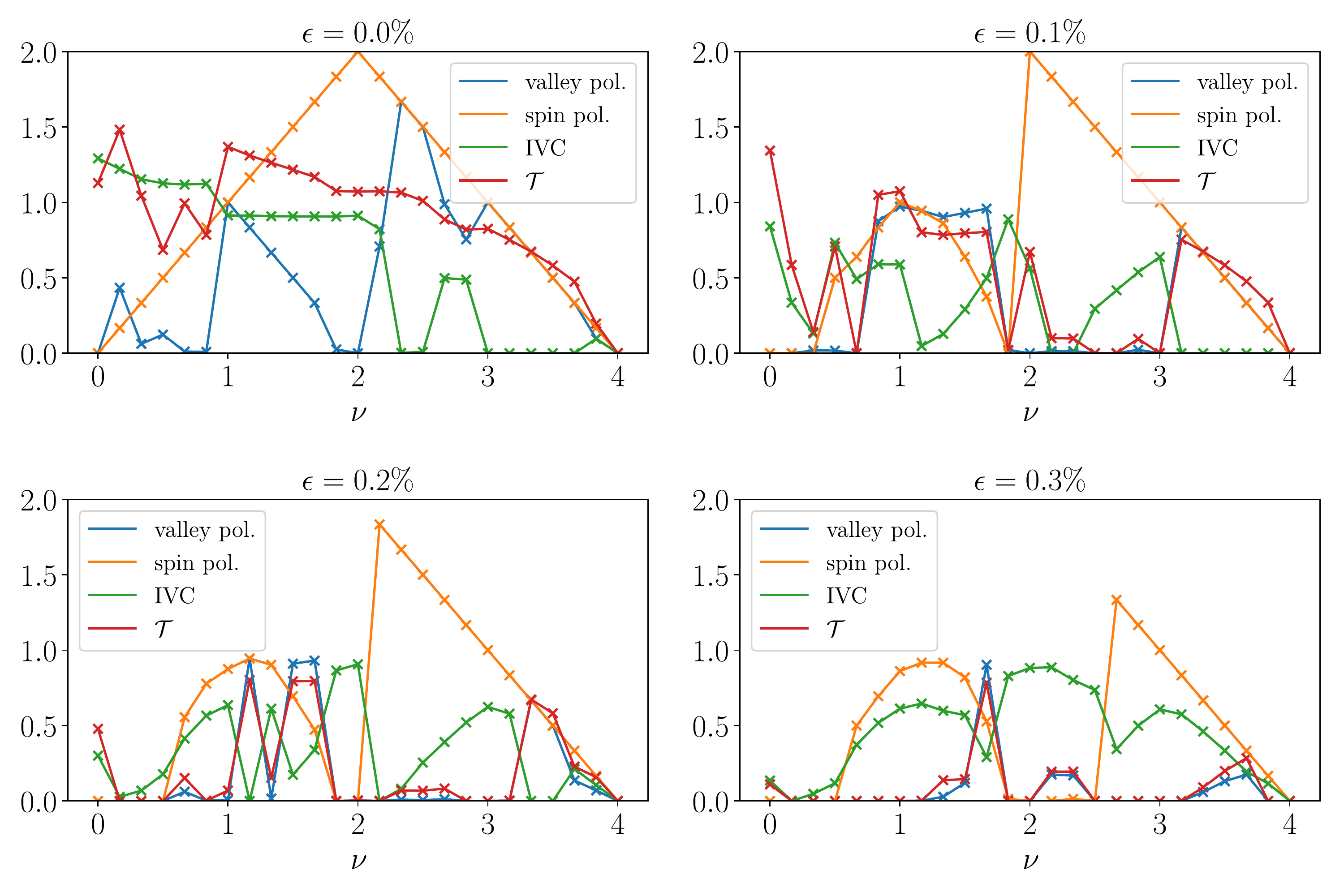}
    \caption{Linecuts at constant strain of the order parameters shown in Fig.~\ref{fig:pols}.}
    \label{fig:linecuts}
\end{figure}

\begin{figure}[h]
    \centering
    \includegraphics[width=0.5\textwidth]{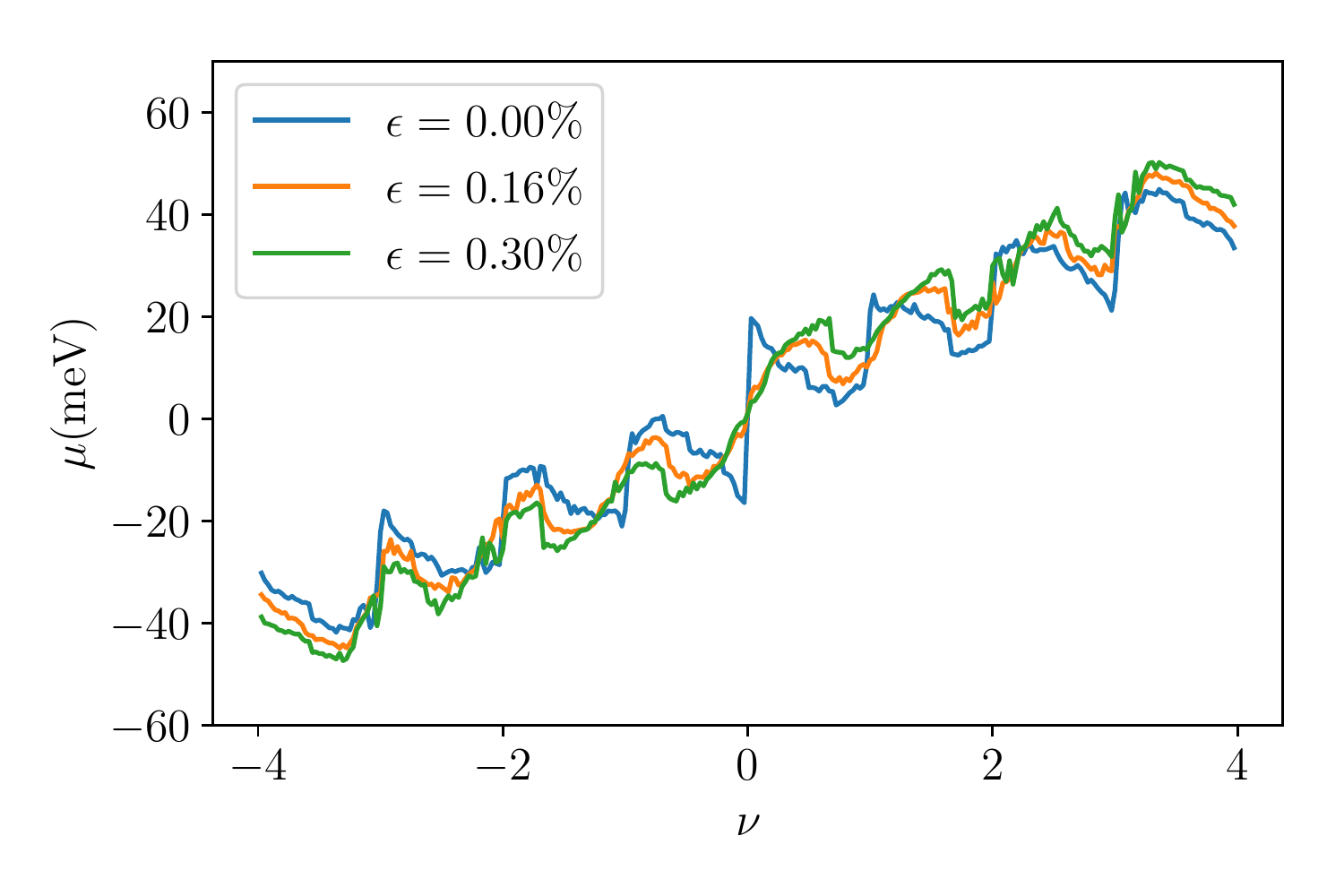}
    \caption{Chemical potential as a function of filling showing the approximate particle-hole symmetry.}
    \label{fig:mu_ph}
\end{figure}

Fig.~\ref{fig:q} shows the magnitude and direction of the IVC wavevector $\mathbf{q}$ of the lowest energy HF solution at a given point in the phase diagram. KIVC order is found at $\mathbf{q}=0$, whereas $\mathcal T$IVC order is found at finite $\mathbf{q}$, i.e.~we have an IKS solution. Fig.~\ref{fig:pols} shows various order parameters characterizing the HF solution. Fig.~\ref{fig:Delta_E} shows the energetics of the solution with and without IVC and compares the energetics of the IKS solutions with different $\mathbf{q}$. Fig.~\ref{fig:C3} plots the order parameter for $\mathcal{C}_3$-breaking. Fig.~\ref{fig:bs} plots the bandstructures at different fillings. Fig.~\ref{fig:finite_T} shows finite temperature HF results. Fig.~\ref{fig:linecuts} presents linecuts of the various order parameters in Fig.~\ref{fig:pols}. Fig.~\ref{fig:mu_ph} shows the chemical potential as a function of filling for both positive and negative filling, demonstrating the approximate particle-hole symmetry.

% \bibliography{bib}
\clearpage

\end{appendix}
\end{document}